\documentclass[12pt]{iopart}

\usepackage{graphicx}
\usepackage{bm}
\usepackage{makeidx}
\usepackage{comment}
\usepackage[ansinew]{inputenc}
\usepackage[usenames,dvipsnames]{pstricks}
\usepackage{subfigure}
\usepackage{epsfig}
\usepackage{amssymb}

\begin{document}
\title[On a deformation of the nonlinear Schr\"odinger equation ]{On a deformation of the nonlinear Schr\"odinger equation}

\author{A Arnaudon$^1$}

\address{$^1$ Department of Mathematics, 
Department of Mathematics, Imperial College, London SW7 2AZ, UK}
\ead{alexis.arnaudon@imperial.ac.uk}

\begin{abstract}
		We study a deformation of the nonlinear Schr\"odinger equation recently derived in the context of deformation of hierarchies of integrable systems. 
		This systematic method also led to known integrable equations such as the Camassa-Holm equation.
		Although this new equation has not been shown to be completely integrable, its solitary wave solutions exhibit typical soliton behaviour, including near elastic collisions.  
		We will first focus on standing wave solutions, which can be smooth or peaked, then, with the help of numerical simulations, we will study solitary waves, their interactions and finally rogue waves in the modulational instability regime. 
		Interestingly the structure of the solution during the collision of solitary waves or during the rogue wave events are sharper and have larger amplitudes than in the classical NLS equation.   
\end{abstract}

\maketitle

\section{Introduction}

In this work, we will study a deformation of the nonlinear Schr\"odinger equation, called hereafter the CH-NLS equation. 
The standard dimensionless form of this equation is given for a complex field $u$ by  
\begin{eqnarray}
		im_t + u_{xx}+ 2\sigma m\left (|u|^2- \alpha^2|u_x|^2\right)=0, \qquad m=u-\alpha^2 u_{xx},
	\label{CH-NLS}
\end{eqnarray}
where $\sigma=\pm 1$ selects between the focussing and defocusing equation. 
Although the two equations have different properties, we will only consider the focussing case, thus we will set $\sigma=1$ for the rest of this paper. \\
This equation was recently derived by the author in \cite{arnaudon2015lagrangian} when developing a theory of a deformation of hierarchies of integrable systems. 
When applied to the AKNS hierarchy of \cite{ablowitz1974inverse} this deformation led to the derivation of several known integrable equations. 
The most important one is the Camassa-Holm equation \cite{camassa1993}, found as a deformation of the KdV equation. 
The CH equation has been extensively studied in the context of shallow water waves, see \cite{camassa1993,dullin2004asymptotically}. 
This has led to the development of the $\alpha$-model, a regularized fluid model which extends the CH equation to higher dimensions (see \cite{holm1998models} and references therein). 
Other deformations of integrable equations derived in \cite{arnaudon2015lagrangian} include the modified CH equation \cite{qiao2006new} and other coupled equations \cite{xia2015new} viewed as deformations of the mKdV and coupled KdV equations respectively. 
These equations all correspond to the deformation of the KdV flow of the AKNS hierarchy.
However, the method of \cite{arnaudon2015lagrangian} can deform the previous flow as well, which corresponds to the NLS equation. 
The CH-NLS \Eref{CH-NLS} is the result of this systematic approach which gathers several integrable equations together in a single framework.  
Although most of the equations derived in this context were independently shown to be completely integrable, this deformation does not guarantee their complete integrability. 
For this reason, there is still no proof for the complete integrability of the CH-NLS \Eref{CH-NLS}. 
The question of the complete integrability of the CH-NLS equation therefore remains an open problem. 

The main focus of this paper we will be on the solutions of this equation, which share the same properties as the solutions of integrable equations. 
We found a one parameter family of standing waves solutions with the remarkable property of having a transition between smooth and peaked shapes by varying this parameter. 
Although very rare in dispersive equations, such peaked solutions have already been found and studied in a generalised NLS equation by \cite{amiranashvili2011dispersion,amiranashvili2014ultrashort}. 
We then numerically studied solitary waves by running initial value problems with hyperbolic secant initial conditions\footnote{Videos of some simulations are available on \url{http://wwwf.imperial.ac.uk/~aa10213}.}. 
We observed the emergence of dispersive and solitary waves. The latter can even have a breathing behaviour, typical of cubic nonlinear equations such as mKdV or NLS for example. 
Numerical simulations of the collisions between these solitary waves show that they keep their identity after a collision, despite numerical errors.
The existence and the stability of solitary waves in non-integrable equations is a vast and difficult subject which will not be addressed here. We refer to \cite{tao2009solitons} for a recent review. \\
It is well known that the existence of solitary waves in most of the nonlinear partial differential equation relies on the balance between nonlinearities and linear dispersions. 
One of the few exceptions is the Camassa-Holm equation \cite{camassa1993} which admits solitons even in the dispersionless case. 
The fundamental mechanism of the CH equation is the nonlinear dispersion, coming from the Helmholtz operator. 
By construction, the CH-NLS \Eref{CH-NLS} has a similar nonlinear dispersion based on the Helmholtz operator but no transformation can remove its linear dispersion, as it is usually used for the study of the CH equation.  
The nonlinear dispersion based on the Helmholtz operator is in fact a common feature of CH type equations, a growing family of integrable equations which admit soliton solutions without linear dispersion (see \cite{camassa1993,degasperis2002new,qiao2006new,hone2008integrable, novikov2009generalizations,xia2015new} among others). 
The CH-NLS equation is thus a very particular equation where the linear dispersion plays a more important role than for other equations of this type. \\
Another interesting fact is that if $m$ is considered as the fundamental field then $u$ is a nonlocal function of $m$ through the relation $u=K\ast m$ with $K=\frac{1}{2\alpha} e^{-|x|/\alpha}$.
The linear dispersion relation is of the form $\omega(\kappa)= \frac{\kappa^2}{1+\alpha^2\kappa^2}$, which is a bounded function. 
Waves of high frequencies with respect to $\alpha$ will thus not propagate.
This effect can be observed in the simulations of collisions where high frequency waves were artificially created by the numerical scheme and then stayed at the same position. 
This type of modifications of the linear dispersion relation for the NLS equation has already been mathematically investigated for example by \cite{colin2009short} in the context of short pulses in optical fibres.
We will also see the CH-NLS equation shows more extreme behaviours than the standard NLS equation, in particular during collisions of solitary waves or rogue waves in the modulational instability regime.  

In \Sref{symmetry} we will look at the conservation laws of the CH-NLS equation as well as a short discussion on its non integrability. 
The peaked and smooth standing wave solutions will be presented in \Sref{standing}, the numerical initial value problems meant to isolate and characterise solitary waves are in \Sref{solitons} and their collision in \Sref{collision}. 
The collisions exhibit more extreme behaviours than the collisions of NLS solitons, namely a higher amplitude and sharper structures during the collisions. 
The same features are found in the modulational instability, studied in the last \Sref{MI} where the CH-NLS Peregrine type solutions are higher and sharper than the NLS Peregrine solutions. 
\ref{numerics} contains the numerical scheme used in this work and a convergence analysis for one of the collision of solitary waves. 

\section{Conservations laws and symmetries}\label{symmetry}
The CH-NLS equation is a Hamiltonian system with Hamiltonian given by 
\begin{eqnarray}
		P = \frac{i}{2}\int (\overline m u_x - m\overline u_x )dx
		\label{P}
\end{eqnarray}
and with the non-canonical Hamiltonian structure of the form 
\begin{eqnarray}
K&=&\left [  
	\begin{array}{cc}
	 -2 m \partial_x^{-1}m & \partial_x + 2m \partial_x^{-1}\overline m\\
	\partial_x + 2\overline m \partial_x^{-1}m& -2\overline m \partial_x^{-1}\overline m
	\end{array}\right].
	\label{K}
\end{eqnarray}
This Hamiltonian structure is exactly the same as second NLS Hamiltonian structure. 
The Hamiltonian $P$ corresponds to the momentum of the solution, as it is associated to the space translation invariance of the equation. 
As will will discuss at the end of this section, the Hamiltonian corresponding to the time invariance of this equation has not yet been found, if it exists. 
A similar feature already exists as noted in \cite{hone2014stability} where the same conserved quantity arises form the space and the time invariance of the equation independently. 

A essential symmetry of the NLS equations is their invariance under phase shifts, which is also shared by the CH-NLS equation.  
For the NLS equation this symmetry is associated with the conservation of the total mass 
\begin{eqnarray}
M=\int |m|^2 dx
	\label{mass}
\end{eqnarray}
which is also a conserved quantity of the CH-NLS equation. 
The associated flow is given by $\left (m_t,\overline m_t\right )^T=J_0\left (\frac{\delta M}{\delta m},\frac{\delta M}{\delta \overline m}\right )^T =\left ( im,-i\overline m \right )^T$, where $J_0:=J(\alpha=0)$ from \Eref{J}.
However, as we will see below, this Hamiltonian structure does not produce the CH-NLS equation and the link between the conservation of mass and the phase shift symmetry seems broken, despite the existence of both properties. 

The CH-NLS equation is invariant under the following ``scaling transformation''
\begin{eqnarray*}
		t\to \lambda^{-2} t,\qquad 	x\to \lambda^{-1} x,\qquad u(t,x)\to \lambda u(\lambda^2t, \lambda x) \quad \mathrm{and} \quad \alpha \to \lambda^{-1} \alpha.
\end{eqnarray*}
This is not a proper transformation as it only relates solutions with different values of $\alpha$.
Nevertheless this scaling can be useful to obtain new solutions and makes clear the importance of $\alpha$ to fix a particular length scale in this equation. 

Another important symmetry of the NLS equation is the Galilean invariance, used to derive the travelling wave solutions from the standing waves. 
This symmetry is absent for the CH-NLS equation, as it is for the CH equation. 
In order to see this fact, one can easily compute the time derivative of the centre of mass $C:= \int x|m|^2dx$ and obtain
\begin{eqnarray}
	\frac{d}{dt}C= i\int  (\overline uu_x-u\overline u_x)dx. 
\end{eqnarray}
The right hand side is not a conserved quantity, thus the evolution of the centre of mass does not follow a simple linear equation in the time variable, as for the NLS equation, where the right hand side would be $2P$.  

\paragraph{On the non-integrability of the CH-NLS equation.} 

First of all, the CH-NLS equation is only known to admit a weak ZCR, as defined in \cite{arnaudon2015lagrangian}. 
This equivalent formulation of \Eref{CH-NLS} cannot guarantee its complete integrability. 
It is actually not yet known if the CH-NLS equation is completely integrable.
We briefly expose here the argument against its complete integrability. 
The possible second Hamiltonian structure which could give a bi-Hamiltonian formulation for the CH-NLS equation would be 
\begin{eqnarray}
	J&=&\left [  
	\begin{array}{cc}
		0 & -i(1-\alpha^2\partial_x^2)  \\
			i(1-\alpha^2\partial_x^2) &0 \\
	\end{array}\right].
	\label{J} 
\end{eqnarray}
Together with $P$ and $M$, they will produce the travelling wave equation $m_t=m_x$. 
Then, the well-known theorem of Magri  would apply to produce the hierarchy of integrable equations and there associated conserved quantities. 
However, this argument fails because the two Hamiltonian structures $J$ and $K$ are not compatible. 
This failure arises in the Jacobi identity of $J+K$ and does not seem to be easy to fix by modification of the Hamiltonian structures\footnote{We are indebted to A.N.W Hone and J.P. Wang for spotting the non compatibility of the two Hamiltonian structures.}. 
We can therefore not predict the existence of other conserved quantities, such as the Hamiltonian itself.

\section{Standing wave solutions}\label{standing}

We derive here a one parameter family of localised standing wave solutions of the CH-NLS \Eref{CH-NLS}.
Although this solution cannot be promoted to a travelling wave solution because of the lack of a Galilean transformation, the standing wave solutions still exhibit interesting properties such as peaked profiles.  
This type of standing waves has already been discovered and investigated by \cite{amiranashvili2011dispersion,amiranashvili2014ultrashort} in the context of short pulses in optical fibres. 

We recall the usual form for standing waves where $\phi(t), a(x)$ are real functions to be determined
\begin{eqnarray}
	u=a(x)e^{i\phi(t)},
\end{eqnarray}
thus
\begin{eqnarray*}
		m=(a(x)-\alpha^2a_{xx}(x))e^{i\phi(t)}\quad \mathrm{and}\quad u_t = i\phi_t u.
\end{eqnarray*}
Inserting this ansatz into \Eref{CH-NLS} and using the separation of variables yield 
\begin{equation*}
		\phi_t= \frac{1}{ a-\alpha^2a_{xx}}\left ( \frac12 a_{xx}	+ (a-\alpha^2a_{xx}) (a^2 - \alpha^2a_x^2) \right). 
\end{equation*}
This allows us to set $\phi=ct$ and integrate the resulting equation after its multiplication by $a_x$ to obtain
\begin{equation*}
	 \alpha^4a_x^4 + (1+2\alpha^2(c -  a^2)) a_x^2	- a^2(2 c - a^2) =0. 
\end{equation*}
We set the integration constant to $0$ as we will only look for solutions vanishing at $\pm \infty$.
The travelling wave ODE can then be derived after calculating the roots of this polynomial for $a_x$ and reads
\begin{equation}
		a_x^2 = \frac{1}{2\alpha^4}\left [ - (1+2\alpha^2( c -  a^2))\pm \sqrt{(1+2\alpha^2 c)^2   -  4\alpha^2a^2}\right].
		\label{stand-ode}
\end{equation}
The height of the stationary wave is given as a function of $c$ by $a_\mathrm{max} =\sqrt{2c}$,  found by setting $a_x=0$ in \Eref{stand-ode}. 
This calculation neglected the fact that there is a square root in the equation. 
Taking it into account gives another limit for the height of the solution and its maximum gradient 
\begin{equation}
	a_\mathrm{max}^\ast = \frac{1+2\alpha^2c}{2\alpha},\quad a_{x,\mathrm{max}}^\ast = \sqrt{c^2-\frac{1}{4\alpha^4}}.	
\end{equation}
We thus have two different standing wave heights which corresponds to two types of solutions. 
If $a_{\mathrm{max}}<a^\ast_\mathrm{max}$ the solution will be smooth, otherwise a peak will appear at the maximum.
Indeed, the condition $\frac{\partial a_x}{\partial a}=0$ is equivalent to having a jump in $a_x$, thus a peaked profile. 
The critical value for $c$ for which the peak appears is $c_\mathrm{crit}=\frac{1}{2\alpha^2}$, and if $\alpha\to0$ then $c_\mathrm{crit}\to \infty$, which is compatible with the non-peaked NLS standing waves. 
\begin{figure}[htpb]
		\centering
		\subfigure[Phase portraits of standing waves.]{\includegraphics[scale=0.59]{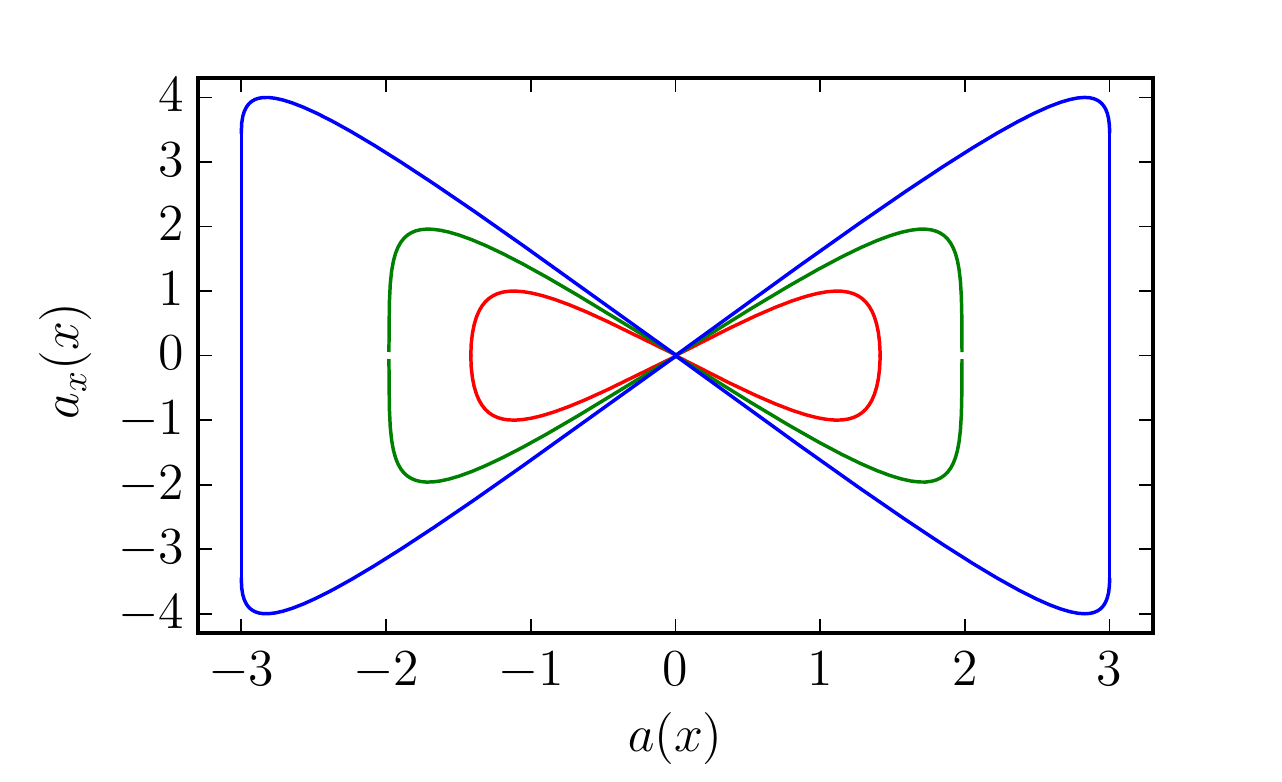}\label{fig:SW-PS}}
		\subfigure[Standing waves]{\includegraphics[scale=0.59]{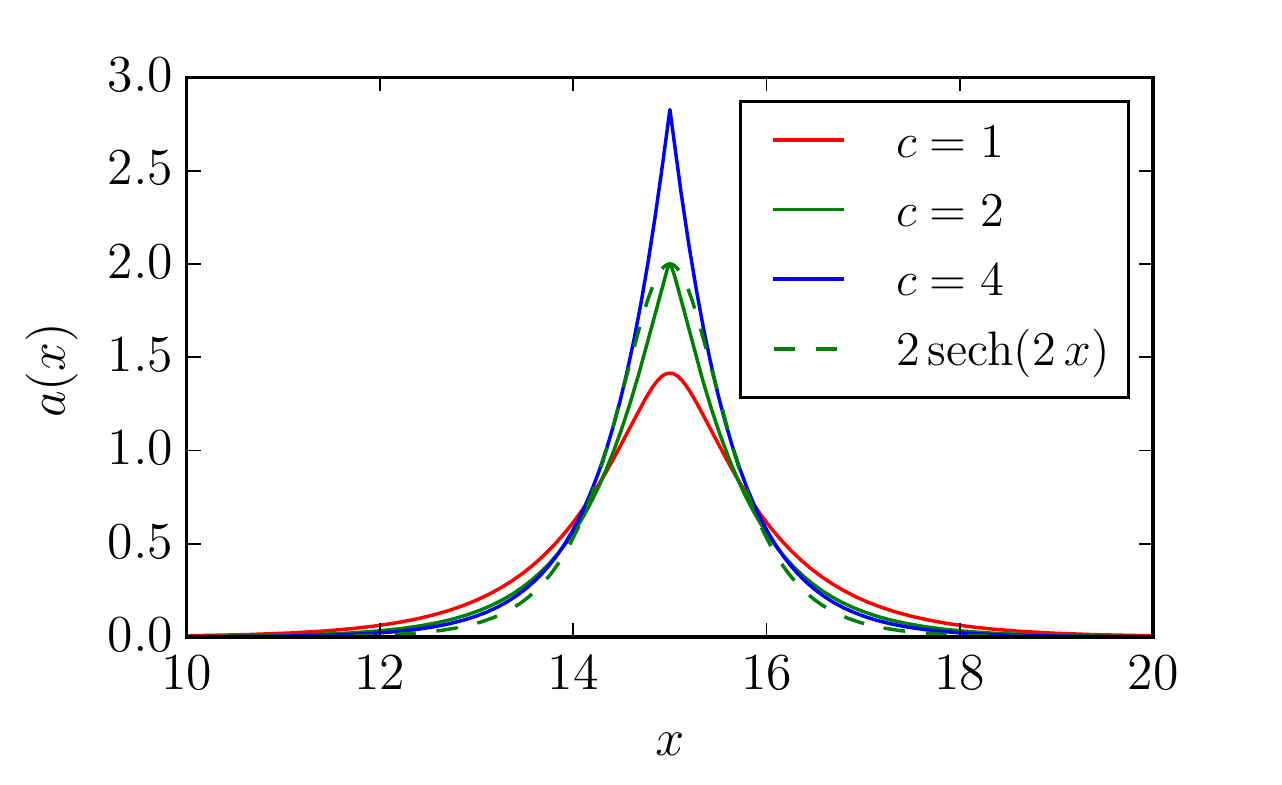}\label{fig:SW-SOL}}
		\caption{The left figure displays the phase portrait for different values of $c$ for $\alpha=0.5$. In this case, the critical value of $c$ for the transition between smooth and peaked profiles is $c_\mathrm{crit}=2$, plotted in magenta. 
		The right figure displays the standing wave profile for the same values of $c$. 
		For large values of $c$, the peak becomes sharper as the jump in the derivative of $a$ becomes larger. 
		As compared to a standard hyperbolic secant profile plotted in dashed line, this standing wave solutions always have sharper peaks and larger tails.  }
\end{figure}
We display on \Fref{fig:SW-PS} the phase portrait of the standing wave solution, given by the right hand side of \Eref{stand-ode} and on \Fref{fig:SW-SOL} the standing wave solution for the same values of $c$. 
We fixed $\alpha=0.5$ and varied $c$ such that the two different solutions appear. 
These solutions have also been tested in the numerics and remain stable, even with the peaked profiles which require higher spacial resolutions. 

\section{Solitary wave solutions}\label{solitons}

The main class of travelling wave solutions that will interest us here are the solitary waves.
In the case of NLS, the solitons have the form 
\begin{eqnarray}
	u(x,t=0)=e^{i\nu x}A\,\mathrm{sech}(A x)
	\label{IC}
\end{eqnarray}
which can then be transformed under a Galilean boost to give the full time dependant solution.
As noted before, this Galilean boost does not exist for the CH-NLS equation which makes this method useless here.
An important point regarding this solution is that $A$ and $\nu$ are two independent free parameters which define a two parameters family of solutions.  
In the CH-NLS equation, the Helmholtz operator will always couple the phase and the amplitude of the solution such that the shape of the solitary wave will depend on the speed.
The definition of $A$ and $\nu$ must therefore be adapted if one want to characterise the CH-NLS travelling wave solutions.
This fact makes the calculation of the travelling waves difficult with the usual travelling wave ansatz and we will therefore only focus here, as a first step, on numerical solutions.
We want to observe solitary waves of the CH-NLS, hence we did some numerical experiments with the hyperbolic secant as initial data. 

\begin{figure}[htpb]
		\centering
		\subfigure[Final solutions of sech initial profiles.]{\includegraphics[scale=0.55]{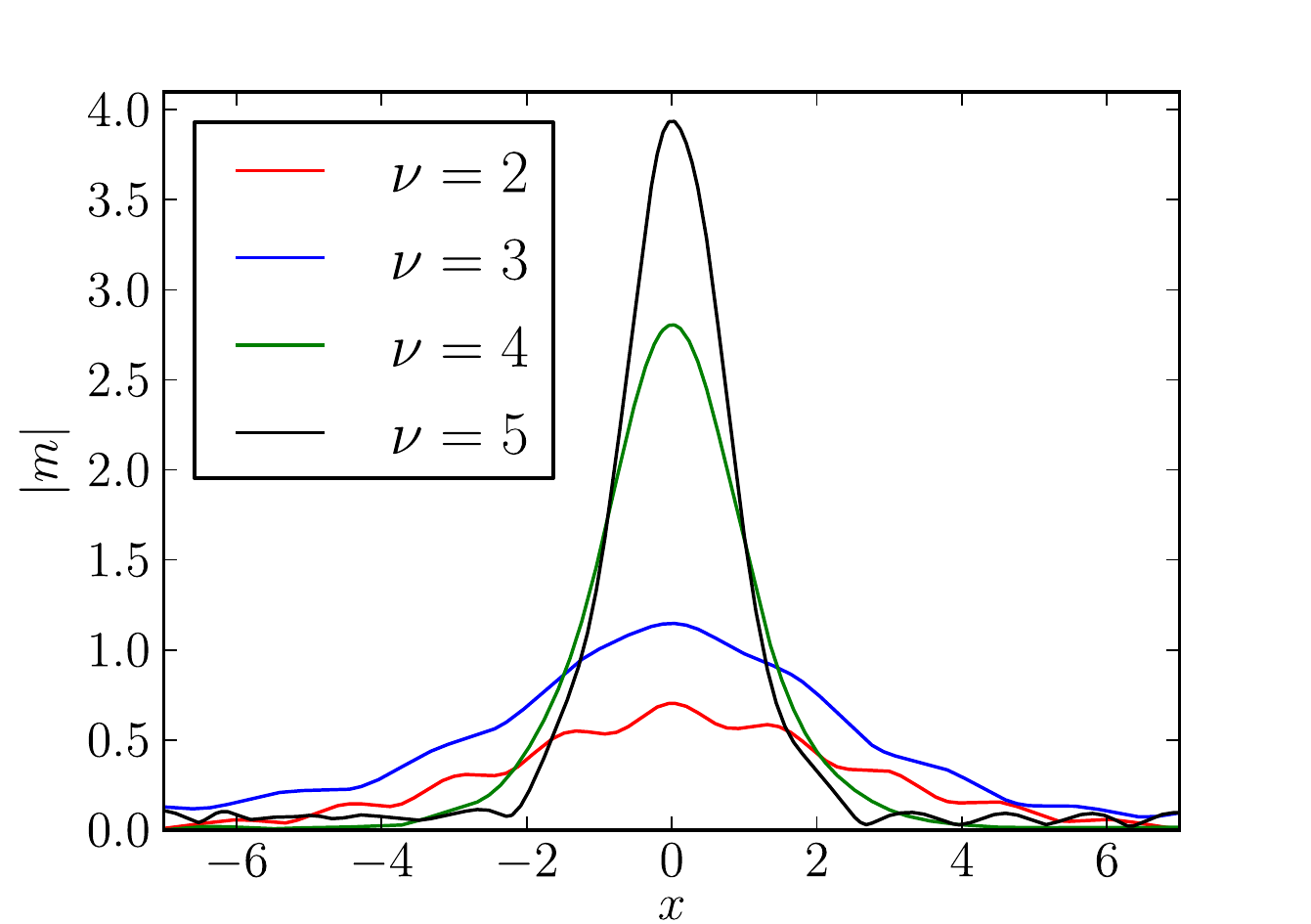}\label{fig:solitons}}
		\subfigure[Space-time plot for $A=1.5$ and $\nu=1$.]{\includegraphics[scale=0.55]{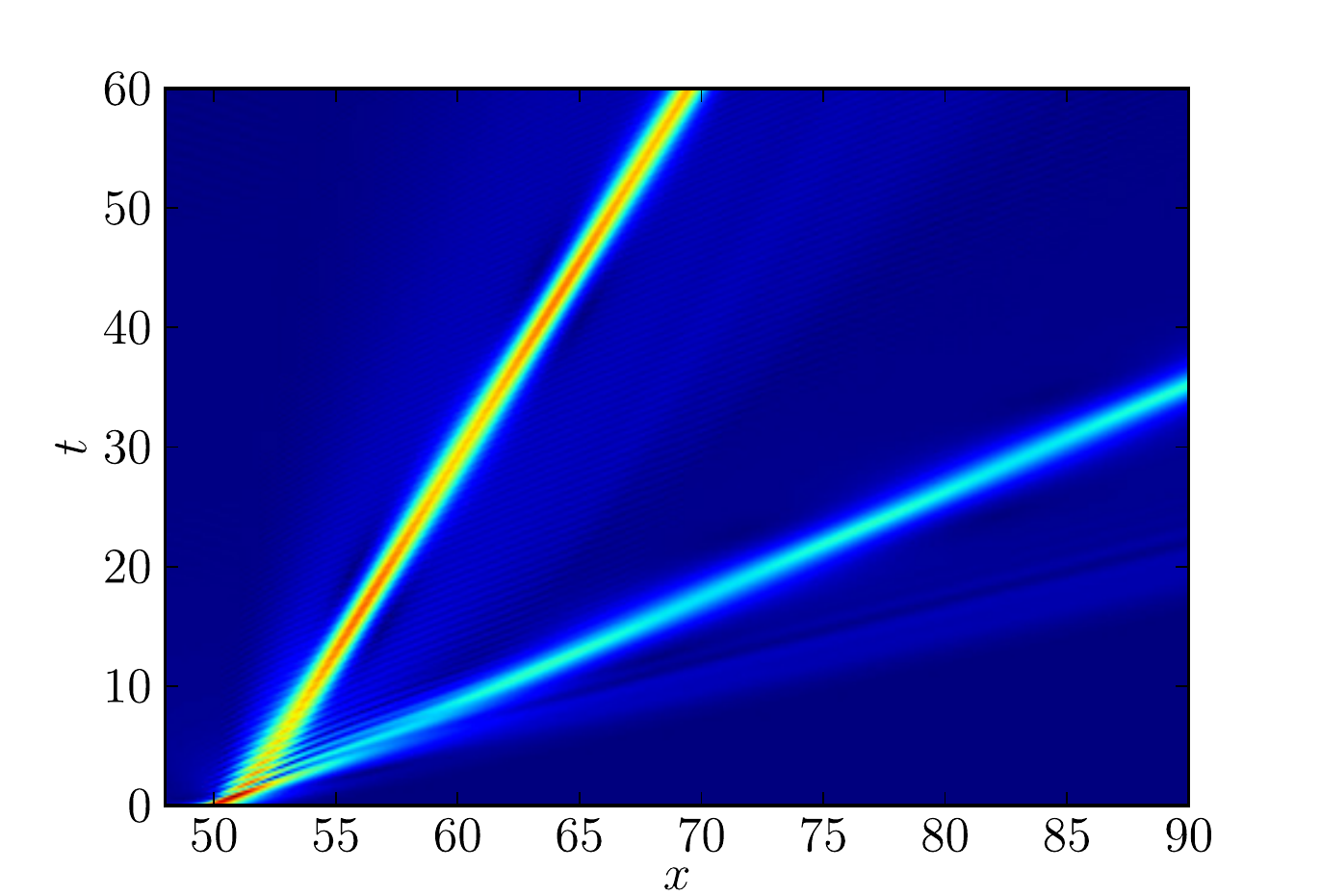}\label{fig:fission}}
		\caption{Numerical solutions of the CH-NLS equation are displayed in both panels. The left panel displays a selection of stable solitary waves. They are not travelling wave as they periodically oscillate in time. 
		The right panel displays the space-time plot of a larger initial condition which undergoes fission and produces two solitary waves as well as dispersive radiations. }
\end{figure}
Recall first that the CH-NLS equation contains a length scale $\alpha$, which can be converted into a scale in wavelength given by $1/\alpha$.  
Provided that the spectral width of the initial data is small compared to $1/\alpha$, namely that $|\kappa_0\pm\Delta\kappa| \ll 1/\alpha$, where $\Delta \kappa$ is the spectral width of the solution and $\kappa_0$ its maximum wavelength, the solution will behave as a NLS soliton. 
In contrast, the higher order terms of the CH-NLS equation will start to play an important role and other type of solutions and behaviour will be observed, such as breathers type solutions.  
Note that in the case of our initial condition in \Eref{IC} we have the simple relations $\Delta \kappa = 1/A$ and $\kappa_0=c$.
The numerical experiments displayed on \Fref{fig:solitons} are snapshots at $t=150$ with  initial profile given by \Eref{IC} with parameters $A=1$ and $\nu=1,2,3,4,5$. 
We also used $\alpha=0.3$ such that the numerical scheme could converge with realistic resolutions.
In each case, one or more solitary wave will emerge from the initial profile as well as dispersive radiations on both sides. 
In order to illustrate this effect, we plotted on \Fref{fig:fission} a simulation from a larger initial pulse with $A=1.5$ which undergoes an initial fission into two solitary waves and dispersive radiations. 
In most of our simulations, the solitary waves at large times are not pure travelling waves, but are of breather type, namely they exhibit periodic motions in time. 
They contain substructures easily identifiable on \Fref{fig:solitons} which periodically oscillate as the solitary wave moves. 
Another interesting fact is that the four solitary waves displayed in this figure have approximately the same phase gradient (roughly corresponding to a $\nu$ in \Eref{IC}), which makes the breathing phenomenon as well as the speed of the solitary wave dependent on the shape of the envelop of the solitary wave and not just on the spectral peak position, as for NLS. 
Indeed, the sharper the solitary wave, the faster.
Although the solutions of \Eref{CH-NLS} are very rich, we will not further pursue this investigation here as we are missing the analytical solutions of the equation and because precise numerically based results about the properties of the solutions are out of the scope of this short study.  

\section{Collision of solitary waves}\label{collision}

Another interesting phenomenon of soliton equations that will be explored in this section is the interaction of solitary waves.  
For integrable equations it is guaranteed that the solitons will keep their identity after the collision but is therefore not sure for the CH-NLS equation. 
Nevertheless we numerically observed collisions with this property, despite some numerical errors (see \ref{numerics}). 
\begin{figure}[htpb]
		\centering
		\subfigure[Slow and large solitary wave.]{\includegraphics[scale=0.5]{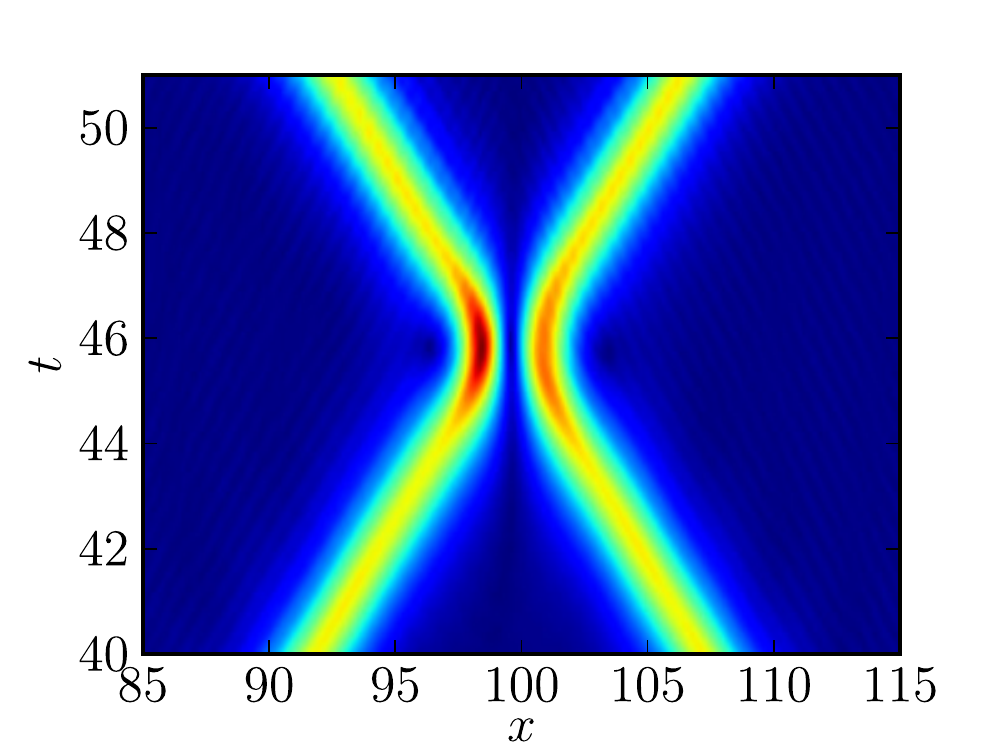}\label{fig:SL}}
		\subfigure[Slow and narrow solitary wave.]{\includegraphics[scale=0.5]{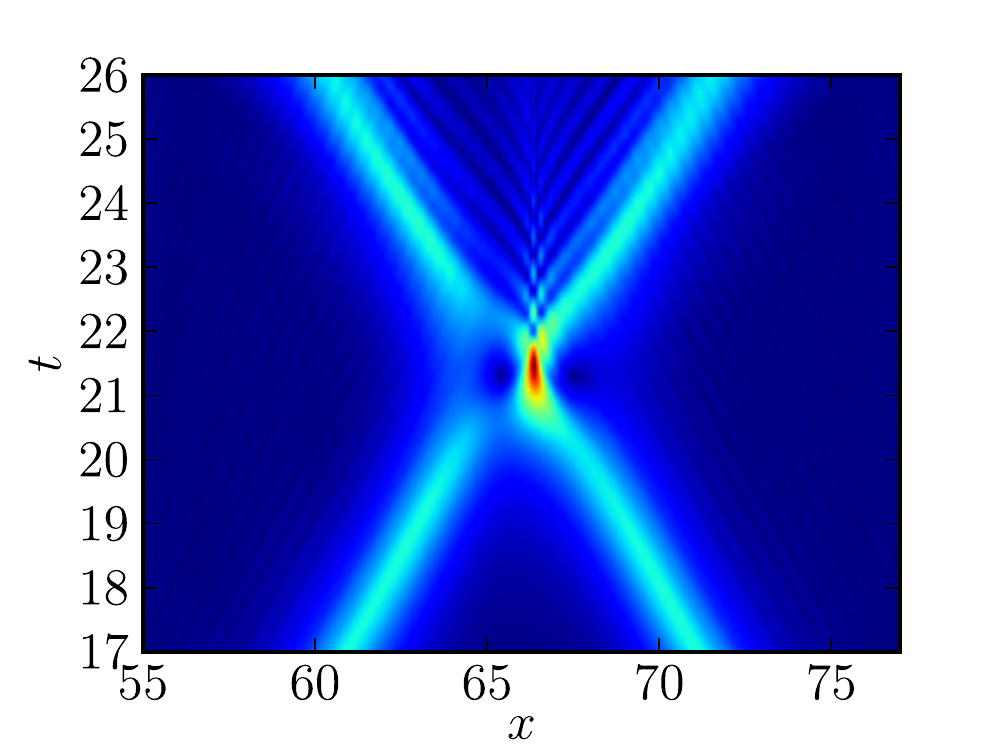}\label{fig:SS}}
		\subfigure[Fast and large solitary wave.]{\includegraphics[scale=0.5]{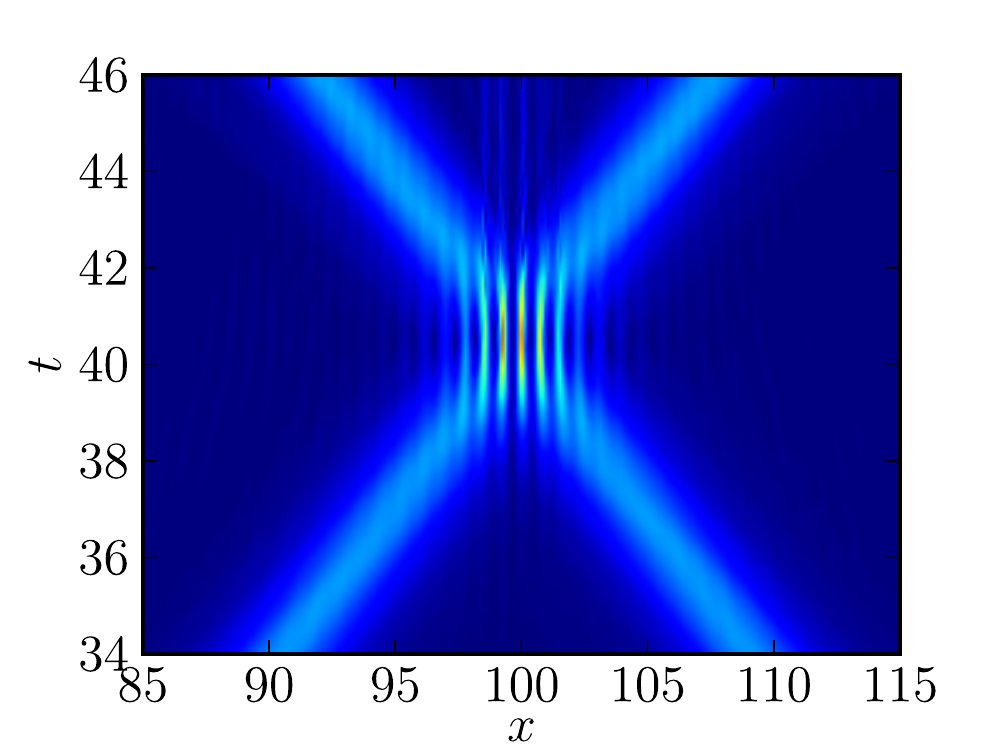}\label{fig:FL}}
		\caption{These figures display three different collisions between two solitary waves of the CH-NLS equation, extracted from initial value problems with sech profiles. 
			The left solitary waves have a narrower spectrum than the other two, hence have a collision more similar to the NLS collision.  
		The other two collisions have higher amplitudes and sharper structures than the first, as one can see from inspection of the differences in colours between the solitary wave and the structures of the collision. 
		This is due to higher wavelengths in the solitary wave spectrum, thus higher nonlinear dispersive effects. 
		The numerical scheme is even not able to fully capture the interaction and high frequency waves remain at the location of the collision. We refer to \ref{numerics} for more details on this numerical error.    }
\end{figure}

We have run three different collisions, with solitary wave extracted from initial value problems as in the previous section.  
We used three different types of solitary wave for this experiment. 
The first solitary waves in \Fref{fig:SL} have a narrow spectrum near the zero wavelength.   
The second solitary waves in \Fref{fig:SS} are sharper than the first, thus have a larger spectrum. 
The last solitary wave in \Fref{fig:FL} have a similar profile than the first, but with a higher speed. 
The first collision is therefore similar to a typical NLS soliton collision whereas the others differ.  
The spectrum of last two type of solitary waves contain higher wavelengths, closer to the scale $1/\alpha$, thus the effect of nonlinear dispersion is more apparent than for the first collision. 
The main difference is that the amplitude and sharpness of the structures created during the collision are bigger than for a standard NLS collision. 
This corresponds to the excitation of higher wavelengths than for NLS equation despite the regularisation feature present in the linear dispersion relation, as mentioned in the introduction. 
This shows that the nonlinear dispersion of the CH-NLS is an important factor for such collision behaviours, mainly used to describe rogue wave events.

\section{Modulational instability}\label{MI}
\begin{figure}[htpb]
		\centering
		\subfigure[Modulational instability gain.]{\includegraphics[scale=0.60]{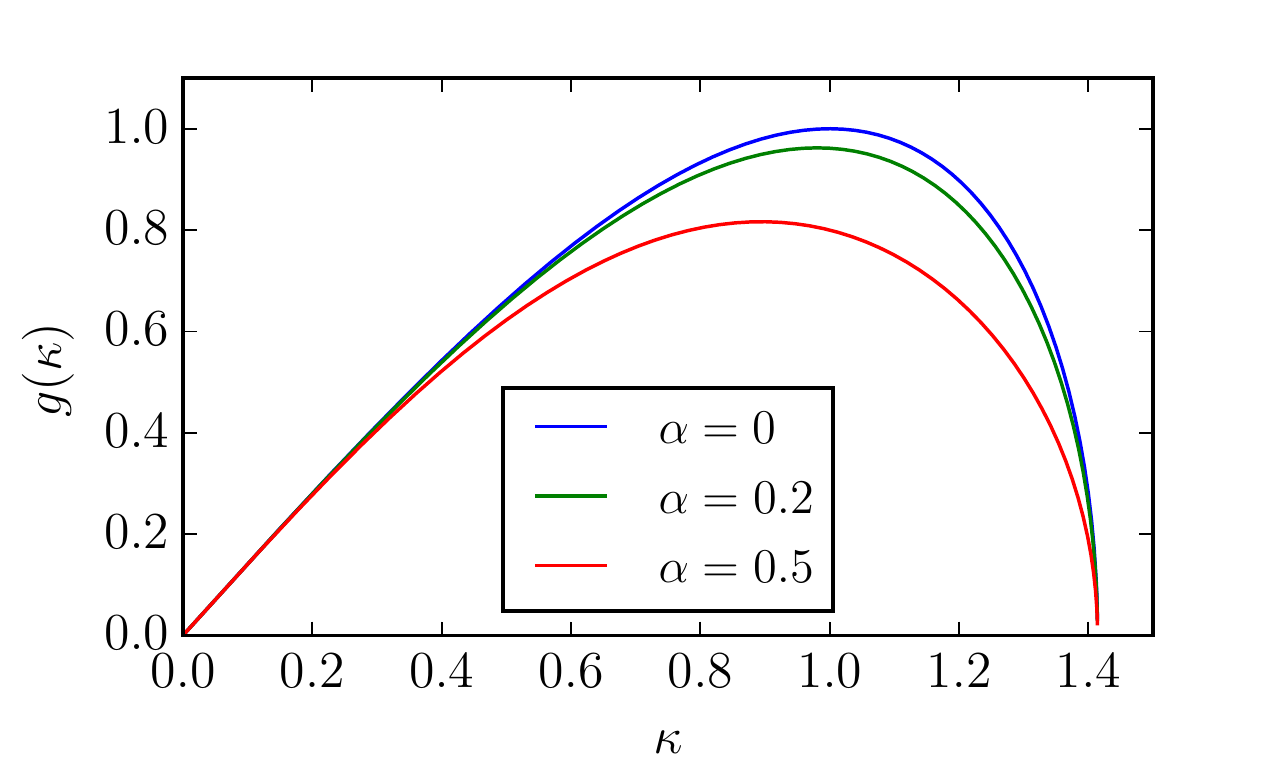}\label{fig:MI}}
		\subfigure[Peregrine solutions.]{\includegraphics[scale=0.60]{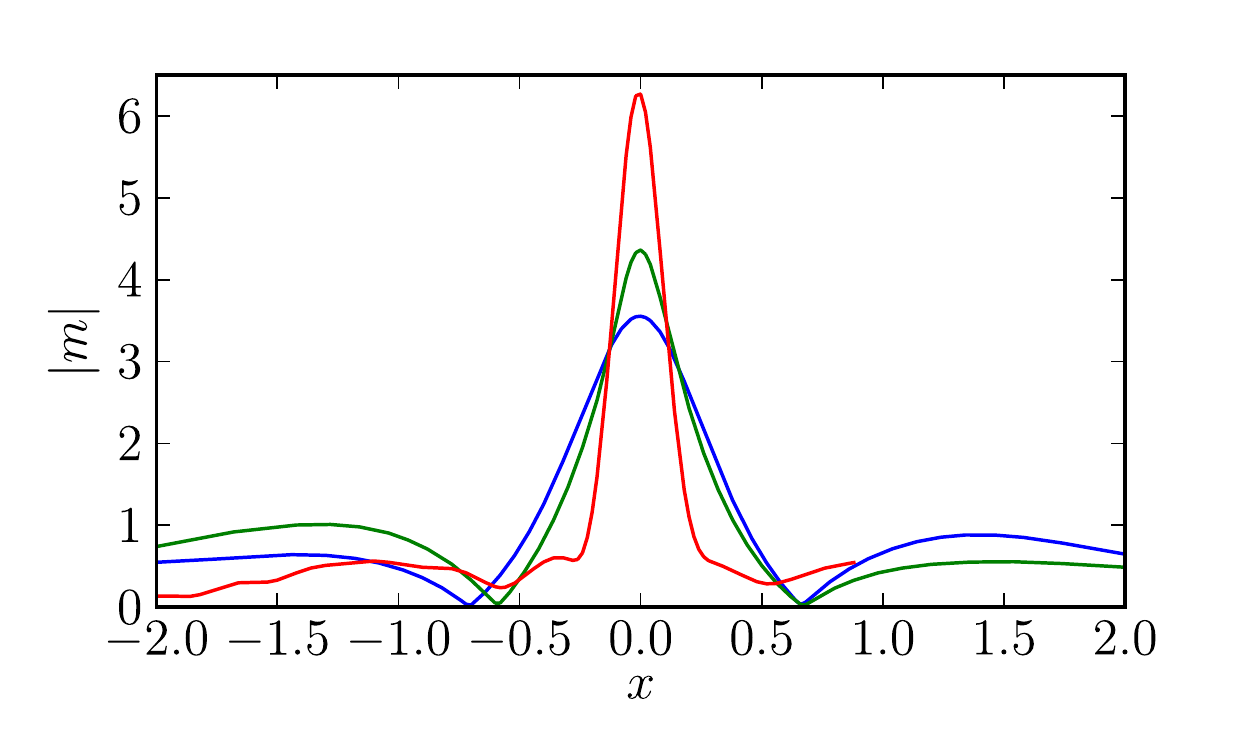}\label{fig:peregrine}}
		\caption{The left panel displays the modulational instability gain of \Eref{MI-gain}, derived from the linearisation of the CH-NLS around the plane wave solution with $P=1$. The gain is reduced but the unstable region not.  
	The right panel exposes a selection of Peregrine type solutions of the CH-NLS equation, extracted from numerical simulation of modulational instabilities with the same colours as for the left panel. 
The nonlinear dispersion of the CH-NLS increased the size of this extreme event, making it interesting rogue wave candidates. }
\end{figure}

We will end this study with some considerations on the modulational instability (MI) \cite{agrawal2007nonlinear,hasegawa1995solitons}, fundamental mechanism of the focussing NLS equation and which is also present in the focussing CH-NLS equation. 
We will follow the standard derivation of the MI gain for the NLS equation.
We first perturb the plane wave solution of \Eref{CH-NLS} with $\epsilon\ll \beta $ as follow
\begin{equation*}
	u(x) = \beta e^{i2 \beta^2 x}\ \ \ \Rightarrow\ \ \ u_\epsilon(x,t) = (\beta+\epsilon(x,t))e^{2i\beta^2x},
\end{equation*}
and then obtain the linearised equation for $\epsilon$
\begin{equation}
	i \epsilon_t - i\alpha^2\epsilon_{xxt} +\epsilon_{xx}+2\beta^2(\epsilon+\overline \epsilon)=0. 
\end{equation}
The growth rate of the MI denoted $g$ can be found using $\epsilon$ of the form $\epsilon =\epsilon_+e^{-i\omega t+i \kappa x}+\epsilon_-e^{i\omega t-i \kappa x}$ and is given by
\begin{equation}
	g(\kappa) = \mathrm{Real}\left (\frac{\kappa\sqrt{ 2\beta ^2- \kappa^2}}{1+\alpha^2\kappa^2}\right). 
	\label{MI-gain}
\end{equation}
The regularisation comes only from the convolution terms of the linear dispersion, hence the MI behaviour will be similar to the NLS equation, but with a smaller growth rate. 
This effect is typical of nonlocal interactions in NLS equations, see for instance \cite{krolikowski2004modulational}.
We plotted on \Fref{fig:MI} the gain $g$ for different values of $\alpha$ with $\beta =1$. 
The effect of the nonlocality only reduces the gain, not the size of the instability region as in \cite{krolikowski2004modulational}.
After that the instabilities have developed, nonlinearities take over and the solution evolves toward a train of pulses. 
It happens that rogue wave events sometimes emerge, with the Peregrine solution being one of the most frequent.
We refer to \cite{dudley2014instabilities} for a recent numerical study of the appearance of rogue waves in the NLS equation. 
We have performed similar experiments with the CH-NLS and extracted candidates for Peregrine solutions at different values of $\alpha$.
All the simulations  we did started with a plane wave of amplitude $1$ and of two periods in a periodic domain of length $50$.
The instabilities developed naturally from the noise induced by the numerical errors.  
We observe a delay in the instability growth, as predicted by \Eref{MI-gain} and the formation of rogue events. 
We displayed the best candidates of Peregrine solution in \Fref{fig:peregrine} for $\alpha=0,0.12,0.18$. 
Similarly to the simulations of collisions of solution, the nonlinear effects are stronger than for the NLS and numerically difficult to capture, hence the low value of $\alpha$ that we could use. 
The shape of the solution are similar and close to the Peregrine solution, but the amplitudes are higher with higher values of $\alpha$. 
The standard factor of $3$ for the ratio of the amplitude of the Peregrine solution with the background amplitude is obviously recovered with $\alpha=0$ and reaches values as large as $6$ with only $\alpha=0.18$.

\section{Summary and open problems}

We presented here our latest results on the CH-NLS \Eref{CH-NLS}, a deformation of the NLS equation, previously derived in \cite{arnaudon2015lagrangian}. 
We first exposed the Hamiltonian structure, the conserved quantities such as the mass and momentum  of this equation and discussed its lack of complete integrability.
Further investigations of the integrability of the CH-NLS are needed, but would have to face technical difficulties, owing to the Helmholtz operator.  \\
We then studied a class of solutions which do not move, the so called standing wave solutions. 
We explicitly found the phase space ODE but only studied the standing waves in the case where the wave vanishes at infinity, thus the non-periodic case. 
The main finding was the existence of a transition form smooth to peaked solutions by varying the only parameter of this class of solutions.  
Unfortunately the lack of Galilean invariance of the CH-NLS equation does not allow the standing wave solutions to be promoted to travelling waves. 
The analytical solution for the travelling wave is actually difficult to compute, hence we performed numerical experiments in order to find and study them.
From initial conditions given by the NLS solitons, we observed the emergence of solitary waves and dispersive radiations. 
Most of the solitary waves exhibit a periodic motion in time, similar to the evolution of a breather of the NLS equation. 
The structure of these oscillations were not studied in details as this would require analytical solutions or more systematic numerical studies. \\
We then studied the collision of solitary waves with the help of numerics and found that the general behaviour is similar to the NLS collisions but with sharper  structures of higher amplitudes.  
This fact is interesting as it is linked to the study of rogue waves, extreme event which are known to appear in nature and are usually described by equations of NLS type. 
Apart from collisions of solitary waves, rogue waves can be found by studying the nonlinear regime of the modulational instability, which is still present in the CH-NLS equation. 
We also found rogue wave events of Peregrine type, but sharper and with higher amplitudes than NLS Peregrine solutions. \\
We finally want to mention that in most of our numerical experiments, the solutions converged well when increasing the space and time resolutions, except at the location of the extreme events such as the collisions of solitary waves or the Peregrine solutions. 
We did a brief analysis of this fact, but we let the development of more accurate numerical schemes for future works. 

\section*{Acknowledgements}
		I want to particularly thanks A. Hone and J. Wang for having spotted a mistake in the proposed proof of complete integrability of the CH-NLS equation and D.D. Holm for helpful comments, suggestions and ideas during all the stages of creation of this paper.
		I am also grateful to J. Elgin, Y. Kodama, B. Xia, Z. Qiao, M. Picasso and T. Ratiu for fruitful and thoughtful discussions during the course of this work. 
		I am thankful to the Imperial College High Performance Computing Service\footnote{ \url{http://www.imperial.ac.uk/admin-services/ict/self-service/research-support/hpc/}} for their HPX facilities where the simulations of this work were run.  
	I gratefully acknowledge partial support from an Imperial College London Roth Award and from the European Research Council Advanced Grant 267382 FCCA.

\section*{References}
\bibliographystyle{iopart-num} 

\bibliography{biblio.bib}
\appendix

\section{Numerical scheme}\label{numerics}
\begin{figure}[htpb]
		\centering
		\subfigure[Numerical error of mass $M$.]{\includegraphics[scale=0.65]{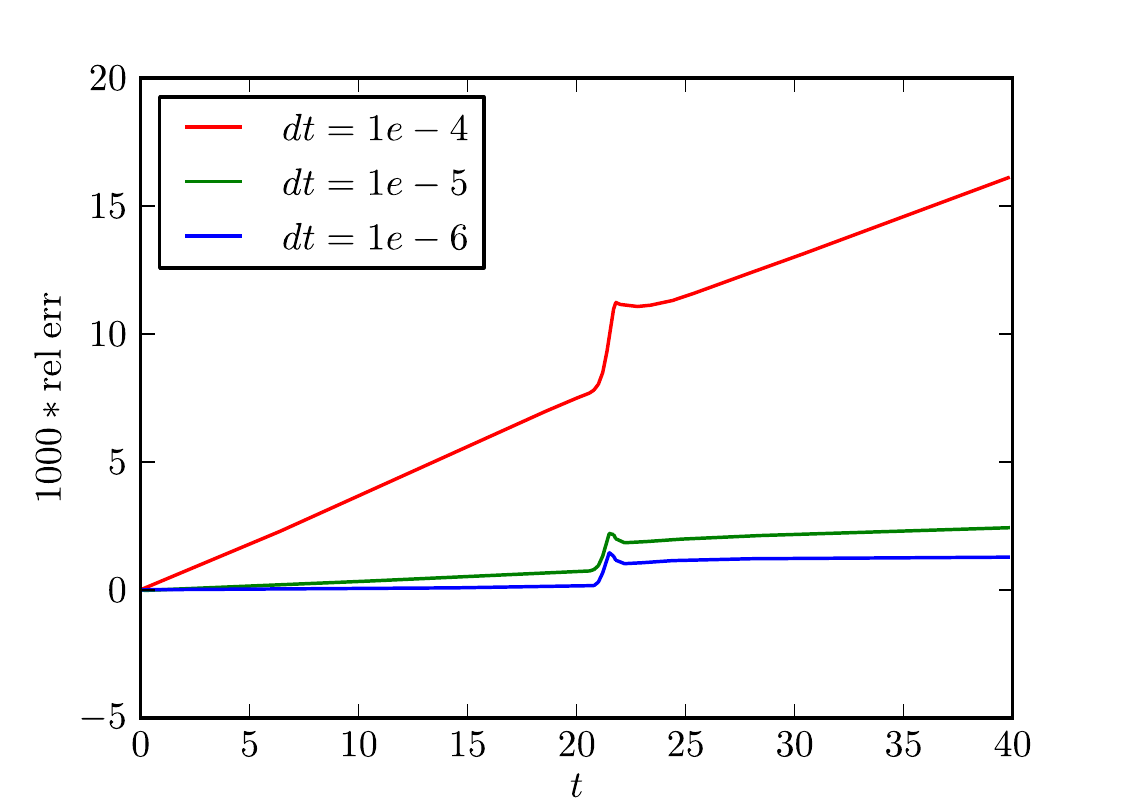}}
		\subfigure[Numerical error of the momentum $P$.]{\includegraphics[scale=0.65]{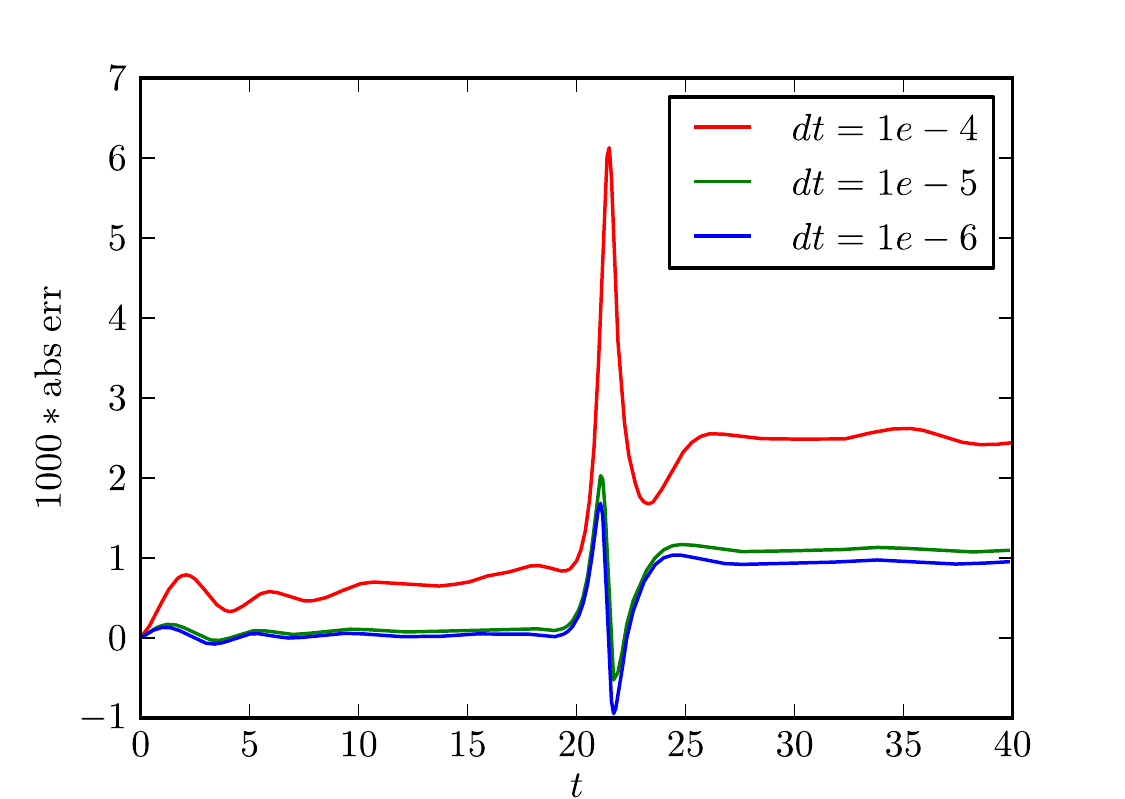}}
		\caption{Both panels of this figure show the errors for the collision displayed in \Fref{fig:SS}. 
	The error for the mass is relative whereas the error for the momentum is absolute, because the value of the momentum is close to $0$. As expected, the convergence is of order $1$, but the collision is never completely resolved, as seen from the jumps in the error of both the momentum and mass. }
		\label{fig:errors}
\end{figure}

The numerical integration of \Eref{CH-NLS} was performed with a pseudo-spectral numerical scheme. 
We used a Fourier spectral method in space, with second order centred finite difference approximation for the space derivatives in the nonlinear term. 
The time integration is second order implicit for the linear part and first order explicit for the nonlinear term. 
The general scheme is given by 
\begin{equation}
		\frac{\widehat u_{i+1} -\widehat u_{i}}{dt} = \frac{i}{1+\alpha^2\kappa^2} \left [-\kappa^2\frac{\widehat u_{i+1}+\widehat  u_i}{2} + \left \{m(|u|^2-\alpha^2|u_x|^2)\right \}_i^{\widehat{\ }}\right]
\end{equation}
where $\widehat u(\kappa)$ is the Fourier transform of $u(x)$ and $i$ runs through the timesteps. 
Although the numerical scheme for the linear term only is not dissipative, the nonlinear term will introduce dissipation.
For each simulation we ensured that the spatial resolution is good enough the capture the full spectrum of the solution. 
We noticed that the CH-NLS equation with large $\alpha$ is numerically difficult to solve, owing to the nonlinear dispersion.
In the simulations of collisions and modulational instabilities we observed the appearance of sharper peaks which eventually breaks into a noisy high frequency set of waves.
These waves do not move, as predicted from the linear dispersion relation, and are purely dispersive.  
In \Fref{fig:errors} we plotted the errors on the conservation of mass and momentum during the collision displayed in \Fref{fig:SS} for three different timesteps. 
This illustrates the first order of convergence of our scheme as well as the difficulties to properly resolve the interactions of solitary waves. 

\end{document}